\documentclass[]{elsart}

\usepackage{times}
\usepackage{epsf}
\usepackage[numbers,sort&compress]{natbib}
\usepackage{amsmath}
\usepackage{amsfonts}
\usepackage{amssymb}
\usepackage{bm}
\usepackage{bbm}
\usepackage{graphicx}
\usepackage{epsfig}
\usepackage{latexsym}
\usepackage{nicefrac}

\usepackage{dcolumn}

\newcommand{\beq}{\begin{equation}}
\newcommand{\eeq}{\end{equation}}
\newcommand{\bea}{\begin{eqnarray}}
\newcommand{\eea}{\end{eqnarray}}
\newcommand{\hf} {\frac{1}{2}}

\def\mr#1{{\mathrm{#1}}}

\def\eq#1{(\ref{#1})}

\begin{document}

\begin{frontmatter}

\title{On the renormalization of the bosonized 
multi-flavor Schwinger model}

\author{I. N\'andori$^{1}$}

\address{
$^1$Institute of Nuclear Research of the Hungarian Academy of Sciences,
H-4001 Debrecen, P.O.Box 51, Hungary}
\date{\today}

\begin{abstract}
The phase structure of the bosonized multi-flavor Schwinger model 
is investigated by means of the differential renormalization group 
(RG) method. In the limit of small fermion mass the linearized RG 
flow is sufficient to determine the low-energy behavior of the 
$N$-flavor model, if it has been rotated by a suitable rotation in 
the internal space. For large fermion mass, the exact RG flow has 
been solved numerically. The low-energy behavior of the multi-flavor 
model is rather different depending on whether $N=1$ or $N >1$, where 
$N$ is the number of flavors. For $N>1$ the reflection symmetry 
always suffers breakdown in both the weak and strong coupling 
regimes, in contrary to the $N=1$ case, where it remains 
unbroken in the strong coupling phase.
\end{abstract}


\end{frontmatter}

{\it Introduction.---} 
Two-dimensional quantum electrodynamics ($\rm{QED_2}$) or the Schwinger 
model \cite{Sc1962} exhibits many analogies with four-dimensional quantum 
chromodynamics (${\rm{QCD_4}}$) including confinement, chiral condensate, 
topological $\theta$-vacuum. The Lagrangian of $\rm{QED_2}$ with massive 
$N$-flavor fermions which is called the $N$-flavor (or multi-flavor) 
Schwinger model reads \cite{CoJaSu1975,Co1976,FiKoSu1979,Ha1976}
\begin{equation}
\label{n_qed_2}
{\cal{L}} = \sum_{n=1}^N {\overline\psi}_n 
\left(i \gamma^{\mu} \partial_{\mu} - m - g \gamma^{\mu}  A_{\mu} \right) 
\psi_n -\frac{1}{4} F_{\mu\nu} F^{\mu\nu}
\end{equation}
where $F_{\mu\nu} = \partial_{\mu}A_{\nu} - \partial_{\nu} A_{\mu}$. 
Using bosonization technique
\cite{Co1975,Co1976,FiKoSu1979,CoJaSu1975,Ha1976,Ha1975,Ge1985} 
the fermionic theory \eq{n_qed_2} can be mapped onto an equivalent Bose form 
\cite{CoJaSu1975,Co1976,FiKoSu1979,Ha1976,Hosotani,GaHiLa1999,FuOn,
Creutz,Sm1997,Ge1985,Lu2000,BySrBuHa2002,ShMu2005,Adam,Af1986,NaSaPo2004} 
\begin{equation}
\label{n_bose_qed_2}
{\cal{L}} = N_{m} \left[
\sum_{n=1}^N \hf (\partial_{\mu} \varphi_n)^2  
+ \frac{\mu^2}{2} \left(\sum_{n=1}^N \varphi_n \right)^2 
- c m^2  \sum_{n=1}^N  
\cos\left(\sqrt{4\pi} \, \varphi_n - \frac{\theta}{N}\right)\right]
\end{equation}
with $\mu^2 = g^2/\pi$, $c = e^{\gamma}/(2\pi)$ where $\gamma = 0.5774$ 
is the Euler's constant, $\theta$ is the vacuum angle parameter,
$N_{m}$ denotes normal-ordering w.r.t. $m$
and $\varphi_n$ $n=1,...,N$ are one-component scalar fields. 
Both the fermionic and the bosonic form of the model has been analyzed 
by various methods from various aspects, e.g. mass perturbation 
theory  \cite{Adam}, density matrix 
renormalization group (RG) method \cite{BySrBuHa2002}, lattice 
calculations \cite{BySrBuHa2002,GaHiLa1999,FuOn}, momentum 
RG method \cite{IcMu1994}, etc. Physical properties (like e.g. chiral 
condensate \cite{Ha1976,Hosotani,Creutz,Sm1997,FuOn}, boson mass 
spectrum \cite{Creutz,GaHiLa1999}) have been investigated for 
arbitrary values of $\theta$, fermion mass and temperature.   

The $N=1$-flavor Schwinger model for $\theta = \pm \pi$ 
has two phases \cite{Co1976,BySrBuHa2002,Lu2000,ShMu2005,Adam}.  
Illustrative and detailed analysis of the phase structure is 
presented in \cite{ShMu2005}. The behavior of the theory is 
controlled by dimensionless ratio $g/m$. For $g/m$ large, i.e. for 
strong coupling, the symmetry $\varphi\leftrightarrow -\varphi$ is 
unbroken, there is a unique vacuum at $\varphi=0$ and there are no 
half-asymptotic particles. For $g/m$ small, i.e. for weak coupling, 
the reflection symmetry suffers spontaneous breakdown, there are two 
vacua approximately located at $\varphi=\pm \sqrt{\pi}/2$ and 
half-asymptotic particles appear. 

The multi-flavor ($N\geq 2$) model has not been studied as extensively 
as the 1-flavor model. However, the relative ignorance toward the 
multi-flavor Schwinger model is perhaps not fully justified as it 
shows more resemblance to ${\rm{QCD_4}}$, because the model features 
a chiral symmetry breakdown. Based on the study of chiral 
condensate \cite{Af1986,Ha1976,Sm1997,Hosotani}, the behavior of 
the Schwinger model has been found to be distinctively different 
for $N=1$ and for $N \geq 2$. Recently, the phase structure of the 
$N=1$ flavor Schwinger model has been investigated by exact functional 
RG method \cite{NaEtAl2007msg} and it has been recovered 
in a rather straightforward way. Our aim in this work is to extend the RG 
analysis for the bosonized multi-flavor ($N\geq 2$) Schwinger model 
to consider its phase structure and clarify the difference 
between the 1-flavor and the multi-flavor models.

The multi-flavor Schwinger model has relevance in solid state physics, 
too. It has been used to describe antiferromagnetic spin chain (see 
e.g. \cite{spin}) and the Bose form of the 1-flavor model has been 
proposed as an adequate model for the description of the vortex 
properties of two-dimensional (2D) isolated thin superconducting 
films \cite{PiVa2000} and the multi-flavor model has been used for
description of vortex dynamics in magnetically coupled layered 
superconductors \cite{NaEtAl2007}. 
The number of flavors of the bosonized multi-flavor Schwinger 
model is equal to the number of layers of the superconducting 
layered system and the Fourier amplitude of the bosonic model 
\eq{n_bose_qed_2} is related to the fugacity of the vortex gas. 
The RG techniques, like the real space RG method 
developed for spin systems usually rely on the dilute gas 
approximation which is equivalent to the linearized RG flow.
However, in order to determine the phase structure and the vortex 
dynamics of layered systems in a reliable manner one has to 
incorporate the effect of the interlayer coupling which requires
corrections to the dilute gas result. Our goal here is to show that 
the the dilute gas approximation, i.e. the linearized RG flow, 
can be used to determine the phase structure of layered  
(multi-flavor) models in the limit of low fugacity (small fermion 
mass) if the original multi-layer (multi-flavor) model has been 
rotated in the internal space. For high fugacities (large fermion 
mass) one has to solve the exact RG flow numerically.

{\it Multi-flavor models.---}
The bosonized multi-flavor Schwinger model \eq{n_bose_qed_2} can be
considered as the specific form of a generalized multi-layer
sine--Gordon (SG) model whose Euclidean action is written as
\begin{equation} 
\label{lsg} 
S = \int {\rm d}^2 r
\left[ \hf (\partial_{\mu} \underline\varphi) 
(\partial_{\mu} \underline\varphi)^{\rm T} +
\hf \underline\varphi \,  {\underline{\underline M}}^2  
\underline\varphi^{\rm T} 
+ \sum_{n=1}^N  y_n \cos (b \varphi_n)
\right] 
\end{equation} 
with the $O(N)$ multiplet $\underline{\varphi}=
\left(\varphi_{1}, \dots, \varphi_{N}\right)$. For the 
specific choice, $b^2=4\pi$,  
and $\underline\varphi \, {\underline{\underline M}}^2 
\underline\varphi^{\rm T} = 
\mu^2 \left(\sum_{n=1}^{N} \varphi_{n}\right)^2$, 
one recovers Eq. \eq{n_bose_qed_2}. The Fourier amplitude
related to the fermion mass ($y\sim m$) and the exact 
relation can be determined by using normal-ordering w.r.t. 
the boson mass. The vacuum angle parameter has
to be chosen as $\theta =\pm N\pi$ for $y_n >0$ and $\theta =0$ 
for $y_n <0$.  In general, SG-type models have also been 
successfully used to investigate vortex dynamics in 2D or 
quasi-2D superconductors \cite{NaEtAl2007,BeCaGi}. Recently, it 
was shown in \cite{NaEtAl2007} that the LSG model with a suitable 
interlayer interaction, 
\begin{equation}
\label{m-lsg}
\hf \underline\varphi \,\,
{\underline{\underline M}}_{\rm{M-LSG}}^2 \,\,  
\underline\varphi^{\rm T} 
= \frac{1}{2} G \left(\sum_{n=1}^{N} a_n \varphi_{n}\right)^2,
\end{equation}
can be used for magnetically coupled layered superconductors
where the coupling strength between the layers denoted by $G$ 
and $a_n=\pm 1$ are free parameters of the model. Based on
symmetry considerations \cite{NaEtAl2007} any choice with 
$a^2_n = 1$ should reproduce exactly the 
same phase structure, as a consequence, the Fourier amplitudes 
(i.e. fugacities) $y_n \equiv y$ for $n=1,2,\ldots,N$. 
The frequency $b^2$ is related inversely to the temperature of 
the corresponding solid-state system. Let us note that, different 
regions of the parameter space have to be considered for the 
condensed matter and and for the high-energy physics problem. 
For the description of the multi-flavor Schwninger model,
one should investigate the phase diagram in the two-dimensional 
plane $y-G$ (for $b^2=4\pi$) and for the vortex dynamics 
one has to consider the phase structure in terms of the frequency 
$b^2$. Let us note, the LSG model with magnetic type 
coupling has a single non-vanishing mass-eigenvalue $M^2_N = N G$.
Another definition for the mass term of Eq.\eq{lsg} 
\begin{equation}
\label{j-lsg}
\hf \underline\varphi \,\,\,\, 
{\underline{\underline M}}_{\rm{J-LSG}}^2 \,\,\,\,  
\underline\varphi^{\rm T} 
= \frac{1}{2} \sum_{n=1}^{N-1}  J (\varphi_{n+1} -\varphi_{n})^2,
\end{equation}
is based on the discretization of the the anisotropic 3D-SG model 
\cite{Na2006} which has been proposed as a suitable model for the 
vortex dynamics of Josephson coupled layered superconductors 
\cite{PiVa1992}. 
Although, it has been shown in \cite{NaJeEtAl2007}, that the LSG 
model with the mass matrix \eq{j-lsg} cannot be used for Josephson 
coupled layered superconductors, 
in order to distinguish between the two types of 
mass matrices, in this paper we refer to \eq{j-lsg} as the 
Josephson-type interlayer interaction. Let us note that 
the Josephson-type LSG model can also be considered as 
a bosonized version of an $N$-flavor fermionic model 
\cite{NaNaSaJe2005,JeNaZJ2006,Na2006}, but not that of the 
multi-flavor Schwinger model \eq{n_qed_2}. In general, the 
LSG model with Josephson type coupling has a single zero and $N-1$ 
non-zero mass-eigenvalues, therefore, the Josephson coupled LSG 
model is invariant under the particular exchange of the layers 
$\varphi_n \leftrightarrow \varphi_{N-n+1}$, hence, 
$y_n \equiv y_{N-n+1}$.

{\it RG approach for multi-flavor models.---}
In this section we summarize briefly the results of the RG analysis 
of LSG type models discussed in our previous publications
\cite{NaNaSaJe2005,NaJeEtAl2007,NaEtAl2007,JeNaZJ2006,Na2006,NaSa2006} 
by means of the approximated form of the Wegner--Houghton \cite{WeHo1973} 
differential RG approach (WH-RG). The WH-RG method provides us the 
complete elimination of the modes in the Wilsonian RG method 
\cite{Wi1971} above the moving momentum scale $k$ which serves as a 
sharp cutoff. In principle any types of RG methods (see e.g. \cite{rg}) 
can be used to consider the behavior of LSG type models. However, 
the usage of sharp momentum cutoff RG is reasonable since a 
spinodal instability \cite{Po2004,tree} may occur during the flow 
\cite{sg2,sg3,NaEtAl2007msg,NaEtAl2007sg,NaSaPo2006} and this can 
be used as a signature of spontaneous breakdown of the symmetry 
$\varphi \leftrightarrow -\varphi$. The WH-RG equation in the 
local potential approximation (LPA) for the LSG 
type models presented in Refs.~\cite{NaNaSaJe2005,NaJeEtAl2007,
NaEtAl2007,JeNaZJ2006,Na2006,NaSa2006} reads as
\begin{equation}
\label{WHdim}
(2+k \, \partial_k) \,\, \tilde V_k ({\underline\varphi}) = 
- \frac{1}{4\pi} \ln \left[ {\mr{det}} \left(\delta_{ij} 
+ \tilde V_k^{ij}({\underline\varphi}) \right) \right],
\end{equation}
where the dimensionless blocked potential ${\tilde V_k} = k^{-2} V_k$
is introduced and $\tilde V_k^{ij}({\underline\varphi})$ denotes
the second derivatives of the potential with respect to $\varphi_i$,
$\varphi_j$. We make the following ansatz for the dimensionless 
blocked potential of the LSG type models
\begin{equation}
\label{def1}
{\tilde V}_{k}({\underline \varphi}) = 
\frac{1}{2} {\underline \varphi} \, \, \, 
{\underline {\underline {\tilde M}}}^{2}(k) \,
{\underline \varphi}^{T}
+ \sum_{n=1}^N {\tilde y_n}(k) \, \cos(b \, \varphi_n)
\end{equation}
where $\tilde y_n(k) = k^{-2} y_n(k)$. Inserting the ansatz (\ref{def1}) 
into Eq.~(\ref{WHdim}), the right-hand side becomes periodic, while 
the left-hand side contains both periodic and non-periodic parts 
\cite{NaNaSaJe2005,Na2006,JeNaZJ2006,NaSa2006}. The non-periodic 
part contains only mass terms, so that we obtain a trivial tree-level 
RG flow equation for the dimensionless mass matrix 
$
\left(2 + k\partial_k \right)
{\underline {\underline {\tilde M}}}^{2}(k) = 0, 
$
which provides the trivial scaling $\tilde J_k = k^{-2} J$ and 
$\tilde G_k = k^{-2} G$, where the dimensionful interlayer couplings 
$J$, $G$ remain constant during the blocking. Finally, we recall that 
in LPA there is no wave-function renormalization, thus the 
parameter $b$ also remains constant during the blocking.
 The argument of the logarithm in Eq. (\ref{WHdim}) must be 
positive. If the argument vanishes or if it changes sign at a 
critical value $k_{\rm{SI}}$, the WH-RG equation (\ref{WHdim}) 
loses its validity for $k<k_{\rm{SI}}$. This is a consequence 
of the spinodal instability (SI) \cite{tree,Po2004}. 
Below the critical scale $k<k_{\rm{SI}}$ the tree-level 
blocking relation (see Eq.(13) of \cite{NaSa2006}) can be used to 
determine the RG flow. In this paper we do not investigate the
tree-level RG flow of LSG type models but we use the appearance 
of the spinodal instability as a signature of the spontaneous
breakdown of the reflection symmetry.

In general, the solution of Eq.\eq{WHdim} can only be obtained  
numerically, however, various approximations of Eq.\eq{WHdim}
are also available in Refs.\cite{NaNaSaJe2005,Na2006,NaSa2006,
JeNaZJ2006,NaJeEtAl2007}. We compare two types of approximations, 
the dilute gas result which is equivalent to the linearized RG 
and the mass-corrected RG flow which incorporates the mass term 
correctly and is able to provide the phase structure 
of the LSG type models in a reliable manner.   
The linearization of the WH--RG equation (\ref{WHdim}) in the full 
potential  around the UV  Gaussian fixed point \cite{ZJ1996} by 
assuming $|\partial_{\varphi_i}^2 {\tilde V_k}|\ll 1$, 
\begin{equation}
\label{WHlin}  
\left(2 + k\partial_k \right) {\tilde V_k({\underline\varphi})} 
\approx \,- \frac{1}{4\pi} \sum_{n=1}^N 
{\tilde V^{nn}_k({\underline\varphi})},
\end{equation}
for the ansatz (\ref{def1}) leads to the linearized WH-RG flow 
\cite{NaSa2006} exhibiting the solutions
\begin{equation}
\label{linsol} 
{\tilde y}_{n}(k) = {\tilde y}_{n}(\Lambda)   
\left(\frac{k}{\Lambda}\right)^{-2 + \frac{b^2}{4\pi}}
\end{equation}
where ${\tilde y}_{n}(\Lambda)$  are the initial (bare) values of 
the fugacities at the high energy ultra-violet (UV) cutoff $\Lambda$.
These are the scaling laws valid at the asymptotically large UV scales 
$(k \sim \Lambda)$, and being independent of the interlayer coupling 
(i.e. mass terms) predicting a phase structure  very similar to  
that of the massless 2D-SG model \cite{sg2,sg_model}. The critical
frequency $b^2_c = 8\pi$ separates the two phases of the model
\cite{Co1975} and the critical temperature is related inversely 
to the critical frequency $T^{\star}_{\rm{KTB}} \sim 1/b_{c}^2 $ 
\cite{sg2,ZJ1996,Po2004}. Let us note that the linearized WH-RG 
equations obtained in LPA for the $N=2$-layer LSG model, are the 
same as those have been found in \cite{PiVa1992} in the dilute gas 
approximation except of the loss of the scale-dependence of $b$ 
due to the usage of the LPA \cite{NaSa2006}. The couplings 
$\tilde J_k$ and $\tilde G_k$ are always a relevant parameters in 
the LSG models and, consequently, the linearization, i.e. the asymptotic 
UV scaling law  (\ref{linsol}), loses its validity with decreasing 
scale $k$ for any value of $b$. 

The simplest way to go beyond the linearized (i.e. the dilute gas) 
approximation and to improve the extrapolating power of the UV scaling 
laws is to take corrections into account of the order ${\cal O}( J/k^2)$ 
for the Josephson and ${\cal O}( G/k^2)$ for the magnetic case, 
which results in the mass-corrected UV scaling laws derived for the 
LSG model in Refs.~\cite{NaNaSaJe2005,Na2006,NaEtAl2007}. This is 
achieved by linearizing the WH-RG equation in the periodic piece of 
the blocked potential, 
\begin{equation}
\label{uv_wh}
(2 + k \, \partial_k) {\tilde U}_k(\varphi_1,..., \varphi_N)
\approx -  \frac{1}{4\pi} \, \,  
\frac{F_1(\tilde U_k)}{C},
\end{equation}
where ${\tilde U}_k(\varphi_1,..., \varphi_N) = 
\sum_{n=1}^N {\tilde y_n}(k) \, \cos(b \, \varphi_n)$ and 
$C$ and $F_1(\tilde U_k)$ stand for the constant and linear 
pieces of the determinant 
$
\det[\delta_{ij} + {\tilde V}^{ij}_k] 
\approx C + F_1(\tilde U_k) + {\cal O}(\tilde U_k^2).
$
Let us first determine the mass-corrected UV scaling laws for the 
LSG model with Josephson type interlayer interaction for $N=2$. 
In this case the solution of Eq.~(\ref{uv_wh}) is 
\cite{NaNaSaJe2005,Na2006}
\begin{equation}
\label{N2}
{\tilde y}(k) =  {\tilde y}(\Lambda) 
\left(\frac{k}{\Lambda}\right)^{\frac{b^2}{8 \pi} - 2}
\left(\frac{k^2 + 2J}{\Lambda^2 + 2J}\right)^{\frac{b^2}{16\pi}}
\end{equation}
with the initial value ${\tilde y}(\Lambda)$ at the UV cutoff 
$k = \Lambda$. From the extrapolation of the UV scaling law 
Eq.~(\ref{N2}) to the IR limit, we can read off the critical 
values $b^2_{c} =16\pi$, for $N=2$. The coupling $\tilde y$ 
is irrelevant for $b^2>b^2_{c}$ and relevant for $b^2<b^2_{c}$. 
The general expressions for the critical frequency and the 
corresponding critical temperature \cite{NaJeEtAl2007} read
\begin{equation}
\label{laydep_j}
b^2_{c} (N) = 8 \pi N, 
\hskip 0.2cm \to \hskip 0.2cm
T^{(N)}_{\rm{J-LSG}} = \frac{2\pi}{b_c^2 (N)} = 
T^{\star}_{\rm{KTB}} \frac{1}{N}
\end{equation}
which are determined previously in a similar manner in the framework 
of the rotated LSG model in Refs. \cite{NaNaSaJe2005,Na2006,JeNaZJ2006}. 
The presence of the coupling $J$ between the layers modifies the 
critical parameter $b^2_c$ of the Josephson coupled LSG model as 
compared to the massless 2D-SG model. 
This important modification can only be deduced if one goes beyond the 
linearized (i.e. dilute gas) approximation, e.g. by the usage of the 
mass-corrected UV scaling laws. 
The similar consideration can be done for the LSG model with magnetic 
type coupling \cite{NaNaSaJe2005,NaEtAl2007}. Since the layers are 
assumed to be equivalent for the magnetically coupled LSG model, 
the RG flow equations for the fugacities of different layers should 
be the same (${\tilde y}_n(k) \equiv {\tilde y}(k)$)  and the 
solution can be obtained analytically
\begin{equation}
{\tilde y}(k) =  {\tilde y}(\Lambda) 
\left(\frac{k}{\Lambda}\right)^{\frac{(N-1) b^2}{N 4 \pi} - 2}
\left(\frac{k^2 + N G}{\Lambda^2 + N G}\right)^{\frac{b^2}{N 8\pi}}
\end{equation}
where ${\tilde y}(\Lambda)$ is the initial value for the fugacity
at the UV cutoff $\Lambda$ and $G$, $b^2$ are scale-independent 
parameters. The critical frequency and the corresponding critical 
temperature which separates the two phases of the model can be read 
directly  
\begin{equation}
\label{laydep_m}
b^2_{c}(N) = \frac{8\pi N}{N-1}, 
\hskip 0.1cm \to \hskip 0.1cm
T^{(N)}_{\rm{M-LSG}} = \frac{2\pi}{b_c^2 (N)} = 
T^{\star}_{\rm{KTB}} \frac{N-1}{N}.
\end{equation}
For $N\to\infty$ the magnetically coupled LSG behaves like a 
massless 2D-SG model with the critical frequency $b_c^2 = 8\pi$.
In principle, one can try to determine the phase structure of 
the LSG models relying on the dilute gas approximation as it 
has been discussed in Ref.\cite{DeGeBl2005} for the 2-layer model. 
However, in this case a 2-stages RG procedure is required. In the 
first step, the real space RG equations are integrated out from the 
UV cutoff ($a_0 \sim 1/\Lambda$) to the effective screening 
length $\lambda_{\rm eff} = 1/\sqrt{2G} = 1/\sqrt{2J}$, where 
the topological defects are taken into account with full flux. 
In the second RG step, from $\lambda_{\rm eff}$ to infinity, 
an a priori assumption has been done by introducing topological 
excitations with fractional flux and, consequently, the 
predicting power of the RG approach has been weakened. In the 
next sections we show that after an appropriate rotation of 
layered models in the internal space, the dilute gas RG 
results can be used to determine the phase structure of
LSG type models without using any a priori assumptions.

{\it RG analysis for rotated models.---}
After performing an $O(N)$ rotation of the layered models which 
diagonalizes the mass matrix, the rotated models do not have 
interlayer interactions, consequently, the rotated fields can 
be treated separately. Let us note that the rotation has generally 
been used for coupled two-dimensional models, e.g. for the SU(N) 
Thirring model \cite{Ha1976} and for the 2-flavor 
\cite{Co1976,FiKoSu1979,GaHiLa1999,FuOn} and for the N-flavor 
Schwinger models \cite{Ge1985,Hosotani,Sm1997,Af1986}. 
The details of the rotation of the $N$-layer Josephson coupled 
LSG model has also been discussed in Refs.~\cite{JeNaZJ2006,Na2006}.
Depending on the number of the non-trivial mass eigenvalues, some 
of the rotated fields have explicit mass terms 
(massive modes) and the other ones are massless, SG-type fields 
\cite{Co1976,FiKoSu1979,Sm1997}. For example, the dimensionless 
potential of the rotated Josephson coupled LSG model contains 
$N-1$ massive fields
\begin{equation}
{\tilde V}_{\rm{J-rot}} = 
\sum_{n=2}^N \frac12 {\tilde M}_n^2 \alpha_n^2 +
\sum_{\sigma_1, ..., \sigma_N} {\tilde w}_{\sigma_1, ...,\sigma_N}
\prod_{n=1}^N e^{i \, \sigma_n \, b_n \, \alpha_n},
\end{equation}
with the rotated $O(N)$ multiplet $\underline{\alpha}^T=
\underline{\underline{O}}^T \, \underline{\varphi}^T$ where
$\underline{\underline{O}}$ represents the rotation with
$b_1^2 = b^2/N$, $b_{n>1}^2 = b^2/(n(n-1))$ and the integer 
valued $\sigma_n$ represent the charges of the topological 
excitations. The rotated magnetic-type LSG model consists of a 
single massive field
\begin{equation}
{\tilde V}_{\rm{M-rot}} = \frac12 {\tilde M}^2 \alpha_1^2 +
\sum_{\sigma_1, ..., \sigma_N} {\tilde w}_{\sigma_1, ...,\sigma_N}
\prod_{n=1}^N e^{i \, \sigma_n \, b_n \, \alpha_n},
\end{equation}
where $M^2 = N G$ and the amplitudes $w_{\sigma_1, ...,\sigma_N}$ 
are different for the Josephson and magnetic LSG models.
At low energies, below the mass-scale the quantum fluctuations are
suppressed by the mass terms producing a trivial scaling for the 
massive modes \cite{ZJ1996}, so, the massive modes can be considered 
perturbatively \cite{JeNaZJ2006,ZJ1996} and they do not influence 
the phase structure of the rotated models. Therefore, one should 
only consider the remaining massless SG fields in order to determine 
the phases of the layered system. At the lowest order of the 
perturbation theory, all the massive modes are set to be equal 
to zero. In this case the effective potential for the rotated 
Josephson type LSG model reads as
\begin{equation}
\label{rot_j_pot}
{\tilde V}_{\rm{J-rot}}(\alpha_1) = 
\sum_{\sigma_1} {\tilde w}_{\sigma_1}
e^{i \, \sigma_1 \, b_1 \, \alpha_1},
\end{equation}
and for the magnetically coupled LSG model the effective potential is 
\begin{equation}
\label{rot_m_pot}
{\tilde V}_{\rm{M-rot}}(\alpha_2, ...,\alpha_N) = 
\sum_{\sigma_2, ..., \sigma_N} {\tilde w}_{\sigma_2, ...,\sigma_N}
\prod_{n=2}^N e^{i \, \sigma_n \, b_n \, \alpha_n}.
\end{equation}
Let us consider the fundamental modes $\sigma_n =\pm 1$. 
The linearized WH--RG equation \eq{WHlin} with 
${\tilde V}_k^{nn} = \partial_{\alpha_n}^2 {\tilde V}_k$, 
for the ansatz \eq{rot_j_pot} and \eq{rot_m_pot} leads to the 
linearized RG flow equations
\begin{align}
(2 + k \partial_k) {\tilde w}_{\rm J}(k) = 
\frac{b_1^2}{4\pi} {\tilde w}_{\rm J}(k),
\hskip 0.3cm 
(2 + k \partial_k) {\tilde w}_{\rm M}(k) = 
\frac{1}{4\pi} \left(\sum_{n=2}^N \, b_n^2 \right)
\, {\tilde w}_{\rm M}(k),
\end{align}
exhibiting the solutions
\begin{align}
{\tilde w}_{\rm J}(k) = {\tilde w}_{\rm J}(\Lambda)   
\left(\frac{k}{\Lambda}\right)^{-2 + \frac{b^2}{N(4\pi)}},
\hskip 0.3cm 
{\tilde w}_{\rm M}(k) = {\tilde w}_{\rm M}(\Lambda)   
\left(\frac{k}{\Lambda}\right)^{-2 + \frac{(N-1)b^2}{N(4\pi)}}
\end{align}
where ${\tilde w}_{\rm J}(\Lambda)$ and ${\tilde w}_{\rm M}(\Lambda)$ 
are the initial values for the Fourier amplitudes at the high energy 
UV cutoff $\Lambda$. The critical value of the frequency parameter 
and the corresponding critical temperatures are found to be
equivalent to Eq.\eq{laydep_j} for the Josephson and Eq.\eq{laydep_m}, 
for the magnetic case. Consequently, the dilute gas RG results for 
the rotated models predict the same layer-dependence of the 
critical temperature as that of obtained by the mass-corrected RG 
for the original LSG-type models.
This proves that the dilute gas approximation is suitable to 
determine the phase structure of rotated layered systems for low 
fugacities. However, for $b^2 < b_c^2$, the
fugacities $\tilde y_n$ are increasing parameters, consequently,
only the exact RG flow is able to determine the phase structure
of the LSG-type models in a reliable manner.

{\it Exact RG flow for the $N=2$ flavor model.---}
Since the results (\ref{laydep_j} and \ref{laydep_m}) have been 
established by an UV 
approximated RG method, it is certainly worthwhile to confirm the 
analysis by a numerical calculation of the exact RG flow. 
Moreover, if one considers the appearance or non-appearance of 
spinodal instability during the blocking which can be used as a
signature of spontaneous symmetry breakdown of the reflection
symmetry, the numerical solution of the exact WH--RG equation is 
required. Indeed, the full RG analysis of the 1-layer LSG model at 
$b^2=4\pi$ discussed in Ref.\cite{NaEtAl2007msg} provides us the 
tool to investigate the symmetric and the symmetry broken
phases of the 1-flavor massive Schwinger model. The critical value 
of the ratio $(\frac{m}{g})_c = 0.311$ which separates the two
phases of the 1-flavor model has been determined by the exact 
RG method which coincides with the results of other calculations 
(see e.g. \cite{BySrBuHa2002}). If $g\gg m$, i.e. below the critical 
ratio, the spinodal instability does not appear during the RG flow, 
therefore, the reflection symmetry remains unbroken. This is the 
consequence of the trivial scaling of the Fourier amplitudes below 
the mass-scale. However, for $g\ll m$, i.e. above the critical value,
the spinodal instability always appears. 
One may assume a similar phase structure for the multi-layer 
model, however, it has been argued in the literature 
\cite{Hosotani,Ge1985,Sm1997} that the low-energy behavior of 
the multi-flavor Schwinger model is different depending on whether 
$N=1$ or $N\geq 2$. Our aim here is to clarify this issue by the 
numerical solution of the exact WH--RG equation derived for the 
2-layer LSG model. We will show that spinodal instability 
always appears for the 2-layer LSG model for $b^2=4\pi$. 

We determine numerically the dimensionless effective potential 
$\tilde V_{\mathrm{eff}}(\varphi_1,\varphi_2)$ for the 
double-layer LSG model as the limit $k\to 0$ of the dimensionless 
blocked potential 
\begin{equation}
\label{n2_ansatz}
{\tilde V}_{k}(\varphi_1,\varphi_2) = 
\frac{1}{2} {\tilde G_k} (\varphi_2 - \varphi_1)^2 
+ {\tilde U}_{k}(\varphi_1,\varphi_2)\,,
\end{equation}
where ${\tilde U}_{k}(\varphi_1,\varphi_2)$ is an arbitrary periodic 
function of the fields (with  $Z_2$ symmetry) including all the 
Fourier modes generated during the RG flow. In order to consider 
the effect of the higher Fourier modes, which were not taken into 
account in the previously utilized linearized and mass-corrected 
linearized WH--RG approach, we use the following ansatz for periodic 
part of the blocked potential
\begin{align}
\label{n2_reduced_ansatz}  
{\tilde U}_{k} = &
\, {\tilde u}_{01}(k) [\cos(b\varphi_1) + \cos(b\varphi_2)] 
+ {\tilde u}_{11}(k) \cos(b\varphi_1)\cos(b\varphi_2) \nonumber \\
& + {\tilde v}_{11}(k) \sin(b \varphi_1) \sin(b \varphi_2),
\end{align}
with the fundamental mode ${\tilde u}_{01} ={\tilde u}_{10} =\tilde y$.
Inserting the ansatz \eq{n2_ansatz} into the exact WH--RG equation
\eq{WHdim} and separating the periodic and non-periodic parts, one 
arrives at the RG equation for the periodic part, see Eq.(17) of 
Ref.\cite{NaSa2006}. For technical reasons, it is more convenient
to consider derivative of the WH--RG equation with respect 
to one of the field variables. By Fourier decomposition, this RG
equation can be reduced to a set of ordinary differential equations 
for the couplings $\tilde u_{01}$, $\tilde u_{11}$ and $\tilde v_{11}$, 
\begin{equation}
\label{rgflow_2LSG} 
{\underline {\underline A}} 
\begin{pmatrix} 
D_k \tilde u_{01} \\[1ex]
D_k \tilde u_{11} \\[1ex]
D_k \tilde v_{11}
\end{pmatrix}
=
\frac{b^2}{4\pi} 
\begin{pmatrix} 
-2(1+\tilde G) \tilde u_{01}  
+ b^2 \tilde u_{01} \tilde u_{11}  \\[1ex]
-2(1+\tilde G) \tilde u_{11} + 2\tilde G \tilde v_{11} 
+ b^2 \tilde u_{11}^2 \\[1ex]
-2(1+\tilde G) \tilde v_{11} + 2\tilde G \tilde u_{11} 
\end{pmatrix}
\end{equation}
where $D_k \equiv (2 + k\partial_k)$ and element of the matrix 
${\underline {\underline A}}$ are 
$
{\underline {\underline A}}_{11} =
-2(1+2\tilde G)+\frac{b^4}{2}(\tilde u_{11}^2 - \tilde v_{11}^2)
$,
$
{\underline {\underline A}}_{22} = {\underline {\underline A}}_{33} = 
-(1+2\tilde G)
$,
$
{\underline {\underline A}}_{12} = {\underline {\underline A}}_{21} = 
b^2(1+\tilde G)\tilde u_{01}-\frac{b^4}{4} \tilde u_{01}\tilde u_{11}
$,
${\underline {\underline A}}_{13} = {\underline {\underline A}}_{31} = 
\frac{b^4}{4} \tilde u_{01}\tilde v_{11},
$ and 
$
{\underline {\underline A}}_{23} = {\underline {\underline A}}_{32} = 0 
$.
We invert the matrix ${\underline {\underline A}}$ and solve the 
RG flow equations for the Fourier amplitudes $\tilde u_{01}$, 
$\tilde u_{11}$ and $\tilde v_{11}$ numerically, by a fourth 
order Runge-Kutta method, whose numerical stability was verified 
by varying the step size. The main advantage of the numerical 
solution of the WH--RG flow equation (\ref{rgflow_2LSG}) is that 
all the non-linear terms are kept. The numerically determined 
scaling of the couplings $\tilde u_{01}(k)$, $\tilde u_{11}(k)$ 
and $\tilde v_{11}(k)$ can be compared to the corresponding 
approximate UV scaling laws. The results are the followings.
%
%
\begin{figure}[ht]
\begin{center}
\includegraphics[width=8cm]{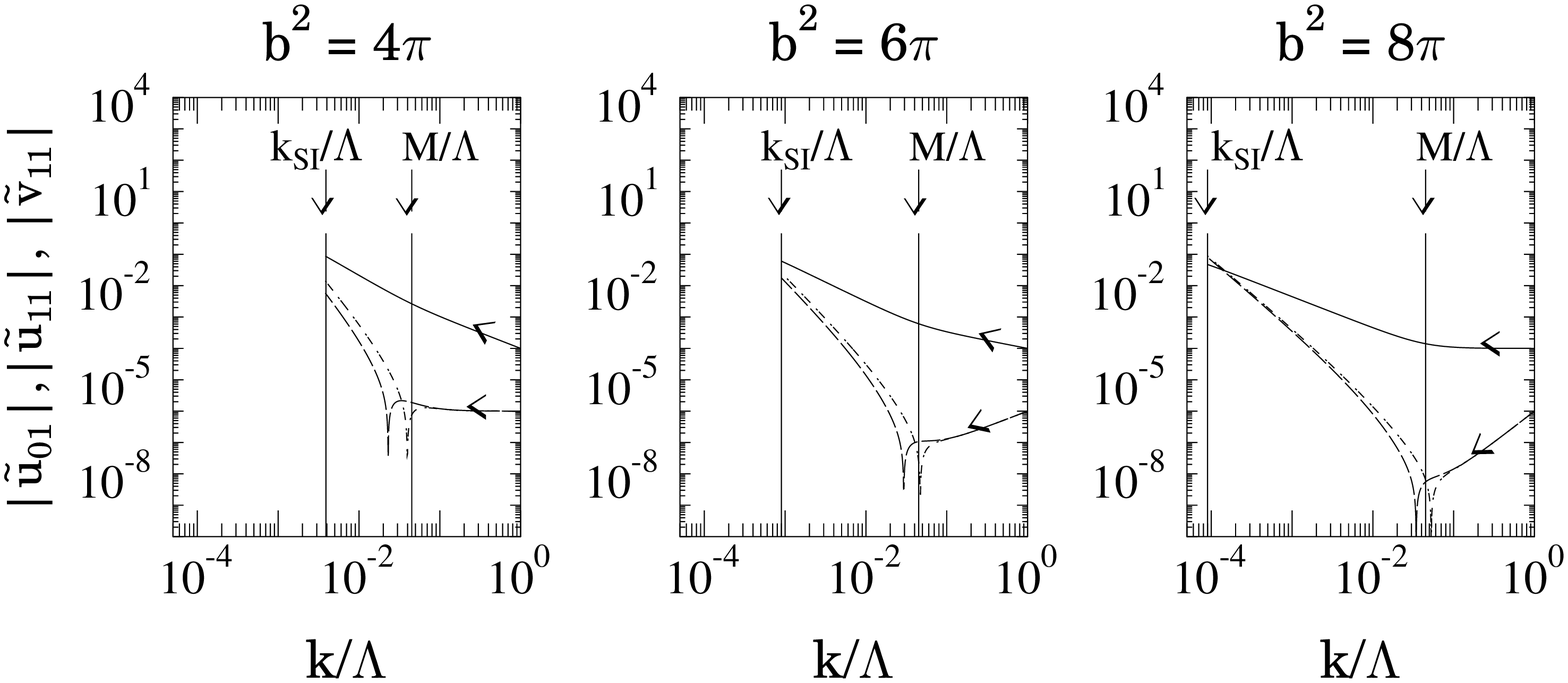}
\caption{\label{fig1} The exact RG scaling of the dimensionless coupling 
constants $\tilde u_{01}$, $\tilde u_{11}$ and $\tilde v_{11}$ of 
the double-layer LSG model is represented graphically for various
frequency parameters. The mass scale is $M/\Lambda = \sqrt{0.002}$. 
The full, dotted and dashed lines correspond to $\tilde u_{01}, 
\tilde u_{11}$ and $\tilde v_{11}$ respectively. Spinodal 
instability ($k_{\rm{SI}}/\Lambda$) appears below the mass scale. 
For increasing value of $b^2$ the momentum scale $k_{\rm{SI}}$
tends to zero and vanishes at $b^2=16\pi$. Below $k_{\rm{SI}}$ 
the WH--RG equation \eq{WHdim} looses its validity and the 
tree-level RG relation (Eq.(13) of \cite{NaSa2006}) has to be used
which is not discussed here.} 
\end{center}
\end{figure}
The mass-corrected UV scaling law \eq{uv_wh} for the fundamental mode
($u_{01}$) coincides with the numerically obtained one. For the 
example, for $b^2 = 12 \pi$, the deviation between the numerical 
solution of the exact WH--RG equation and the solution of the 
mass-corrected UV linearized WH--RG equation at the scale 
$k= 1.0 \times 10^{-5}$ is $9.69\times 10^{-7}$. This 
coincidence demonstrates that the flow of the fundamental 
coupling $\tilde u_{01}$ is well described by 
the mass-corrected UV scaling law if no spinodal instability 
appears during the blocking. Therefore, we find that for 
low fugacities (small fermion mass), the phase structure of the 
double-layer LSG model obtained numerically is the same as that 
predicted by the extrapolation of the mass-corrected UV scaling 
laws, because the higher harmonics do not modify the scaling of 
the fundamental mode (see Figs.~\ref{fig1} and \ref{fig2}).
%
%
\begin{figure}[ht]
\begin{center}
\includegraphics[width=8cm]{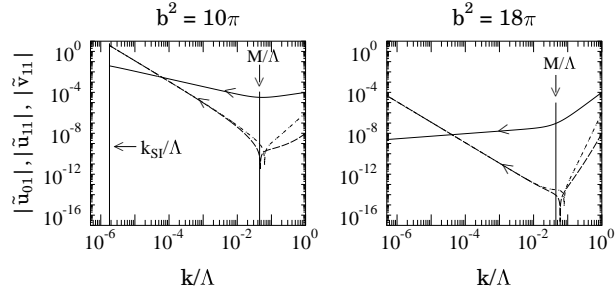}
\caption{\label{fig2} The scaling of the dimensionless coupling 
constants of the double-layer LSG model is shown for two different 
frequencies $b^2 = 10 \pi$ and $b^2 = 18 \pi$ and for the mass 
scale $M/\Lambda = \sqrt{0.002}$. The UV ($M \ll k$) and IR 
($k \ll M$) regions are separated by the mass eigenvalue. 
The scaling of the fundamental mode $\tilde u_{01}$ (full line) 
depends on the frequency, increasing (relevant) for $b^2<16\pi$ 
and decreasing (irrelevant) for $b^2>16\pi$. The initial values 
and consequently the UV scalings are different for $\tilde u_{11}$ 
(dotted) and $\tilde v_{11}$ (dashed) but in the IR region the 
trajectories are coincides resulting in a trivial tree-level 
scaling ($\sim k^{-2}$) which is independent of the frequency.
}
\end{center}
\end{figure}
The IR behavior of the higher harmonics and 
consequently the effective potential can only be determined 
by the numerical solution of the exact WH--RG equation when 
all the non-linear terms are kept. For example, the UV 
approximated RG flow is not able to determine the sign changes 
(see the peaks of the dotted and dashes lines in the figures) 
of the higher harmonics. According to the exact RG flow, 
in the IR limit $u_{11}$ and $v_{11}$ coincide independently 
of their initial values at the UV cutoff (see Fig.~\ref{fig2}). 
Therefore, the low-energy effective potential of the 2-layer 
LSG model determined by the exact RG approach is 
\begin{equation}
\label{eff_pot}
{\tilde V}_{\rm{eff}} = 
\frac{\tilde G}{2} (\varphi_2 - \varphi_1)^2 
+ {\tilde u}_{01} [\cos(b\varphi_1) + \cos(b\varphi_2)]
+ {\tilde w}_{11} \cos(b\varphi_1 -b\varphi_2), 
\end{equation}
with ${\tilde w}_{11}\equiv {\tilde u}_{11}={\tilde v}_{11}$
which corresponds to the massive mode in the rotated form
of the model, hence, $w_{11}$ has a trivial tree-level IR 
scaling ($\sim k^{-2}$) which is independent of the frequency 
$b^2$.
The numerical solution of the exact WH--RG equation
shows (Fig.~\ref{fig1}) that for the 2-layer LSG model with 
$b^2=4\pi$, the spinodal instability (SI) always appears during the 
RG flow in contrary to the 1-layer model where the appearance 
of SI can be avoided for sufficiently small initial value for 
the fugacity. For the 1-flavor model for low fugacity, the SI
cannot be detected below the mass scale since the presence
of the mass term predicts a trivial tree-level scaling for 
all the Fourier amplitudes. For the 2-flavor model, below 
the mass scale, only the massive modes have trivial scalings 
but not for the fundamental one ($u_{01}$). Hence, the 
appearance of SI is unavoidable.

{\it Summary.---}
In this paper we investigated the phase structure and the low-energy 
behavior of the bosonized multi-flavor Schwinger model 
\eq{n_bose_qed_2} by means of the Wegner--Houghton RG method. 
The Bose form of the multi-flavor model consists of 2D
sine--Gordon fields coupled by an appropriate mass matrix \eq{m-lsg} 
which has also been used to describe the vortex properties 
of magnetically coupled layered superconductors \cite{NaEtAl2007}. 
Another definition (see Eq.\eq{j-lsg}) for the mass matrix of 
the multi-flavor i.e. layered sine--Gordon model has also been 
discussed which is based on discretization of the 3D
sine--Gordon model \cite{Na2006}.
Let us note, that the layered model with the mass matrix \eq{j-lsg} 
can in principle also be considered as a Bose form of a fermionic 
model as it has been argued in \cite{Na2006,JeNaZJ2006}, 
however not that of the multi-flavor Schwinger model \eq{n_qed_2}. 

It has been shown that in the limit of small fermion mass the 
linearized RG flow (i.e. the dilute gas approximation) is sufficient 
to determine the phase structure of the multi-flavor model in a 
reliable manner, if it has been rotated  by a suitable  rotation
in the internal space which diagonalizes the mass matrix. This 
receives important application in condensed matter physics where 
the usual RG techniques are based on the dilute gas approximation.
For example, using the idea of rotation the vortex dynamics 
of magnetically coupled layered superconductors can be considered 
by means of dilute gas RG methods and no two-stages RG 
\cite{DeGeBl2005} is required.
The calculation of the exact Wegner--Houghton RG flow of the 
2-flavor (i.e. 2-layer) layered sine--Gordon model, by a numerical 
approach including higher-order Fourier modes confirms that the 
mass-corrected UV scaling law (i.e. the linearized RG flow for the 
rotated model) is sufficient to determine the phase structure of 
the layered sine--Gordon model in the low fugacity (small fermion 
mass) limit. 

The rigorous RG study of the layered sine--Gordon model verifies 
that the low energy effective potential of the bosonized 
multi-flavor Schwinger model is a sine--Gordon type model which 
undergoes a KTB-type phase transition at the flavor-number 
dependent critical frequency $b_{c}^{2}(N) = 8\pi N/(N-1)$. 
In the limit $N\to\infty$, the layered sine--Gordon model 
tends to the two-dimensional sine--Gordon theory with the 
critical frequency $b_c^2 =8\pi$. 
Therefore, in the large $N$ limit, the low energy behavior
of the bosonized multi-flavor Schwinger model becomes 
independent of the boson mass i.e. the coupling $g$.
This is consistent with the flavor-dependence of the chiral 
condensate $<\overline{\psi},\psi> \sim m^{(N-1)/(N+1)} g^{2/(N+1)}$, 
and the mass gap $M_{\rm{gap}}\sim m^{N/(N+1)} g^{1/(N+1)}$ 
which are independent of $g$ if $N\to\infty$ \cite{Sm1997}.
The numerical solution of the exact RG equation shows that 
in case of the 2-flavor (i.e. 2-layer) layered sine--Gordon 
model with $b^2=4\pi$, the spinodal instability always appear 
during the RG flow in contrary to the 1-layer model where the 
appearance of spinodal instability can be avoided for 
sufficiently small initial value for the fugacity.  Consequently,
for $N>1$ the reflection symmetry always suffers breakdown in 
both the weak and strong coupling regimes, in contrary to the 
$N=1$ case, where it remains unbroken in the strong coupling 
phase.

Finally, let us mention that the extension of the RG analysis 
presented in this paper can also be used to consider the 
$\theta$-dependence of the multi-flavor model and to map out the 
phase structure of the multi-frequency sine--Gordon type models 
\cite{double_sg}.

{\it Acknowledgment.---}
The author acknowledges the numerous fruitful discussions with 
U. D. Jentschura, S. Nagy, K. Sailer, K. Vad and the warm hospitality 
during a visit to the Max--Planck--Institute for Nuclear Physics 
(Heidelberg).

\end{document}